\documentclass[12pt,preprint]{aastex}

\shorttitle{Long Term Light Curve of GM Cep}
\shortauthors{Xiao, Kroll \& Henden}

\begin{document}


\title{Long-Term Light Curve of Highly-Variable Protostellar Star GM Cep}


\author{Limin Xiao}
\affil{Physics and Astronomy, Louisiana State University, Baton Rouge, LA, 70803}

\author{Peter Kroll}
\affil{Sonneberg Observatory, Sternwartestr. 32, D - 96515 Sonneberg, Germany }

\author{Arne A. Henden}
\affil{American Association of Variable Star Observers, 49 Bay State Rd., Cambridge, MA 02138}



\begin{abstract}

We present data from the archival plates at Harvard College Observatory and Sonneberg Observatory showing the field of the solar type pre-main sequence star GM Cep. A total of 186 magnitudes of GM Cep have been measured on these archival plates, with 176 in blue sensitivity, 6 in visible, and 4 in red. We combine our data with data from the literature and from the American Association of Variable Star Observers to depict the long-term light curves of GM Cep in both B and V wavelengths. The light curves span from 1895 until now, with two densely sampled regions (1935 to 1945 in B band, and 2006 until now in V band). The long-term light curves do not show any fast rise behavior as predicted by an accretion mechanism. Both the light curves and the magnitude histograms confirm the conclusion that the light curves are dominated by dips (possibly from extinction) superposed on some quiescence state, instead of outbursts caused by accretion flares. 
Our result excludes the possibility of GM Cep being a FUor, EXor, or McNeil's Nebula type star. Several special cases of T Tauri stars were checked, but none of these light curves are compatible with that of GM Cep. The lack of periodicity in the light curve excludes the possibility of GM Cep being a KH 15D system. 

\end{abstract} 

\section{Introduction}

GM Cep is a solar type variable star in the $\sim$ 4 Myr-old open cluster Tr 37 \citep{sic04, sic05}, which is located at a distance of 900 pc \citep{con02}. The coordinates of GM Cep are $21^h 38^m 16^s.48$ and $+57^{\circ} 32'  47''.6$.  It has a late-type spectral classification of G7V-K0V, with a mass of $\sim$2.1 $M_\sun$ and radius estimate of 3 - 6 $R_\sun$ \citep{sic08}. A companion star has been hypothesized as part of the physical mechanism for the variability in the GM Cep system, but it has not been seen. 

The first recorded photometric data for GM Cep was taken at Sonneberg Observatory \citep{mor39} and showed with the visual magnitude varying from 13.5 to 15.5 mag. \citet{suy75} showed that GM Cep had a stable period of up to $\sim$100 days, and it was experiencing rapid variation between 14.2 and 16.4 mag. \citet{sic08} listed and summarized the available data in the literature and depicted a long term light curve in multiple wavelengths. This list contains 16 magnitude, in V band and 5 in B band, most of which were taken in 2006 and later. The only one B band magnitude before 2006 was taken from \citet{kun86}, with B = 17.31 mag. It is much fainter than any other available B magnitude values, and the simultaneous V magnitude is not significantly high. \citet{sic08} took it as an outlier and did not include it in their analysis. Although the data for GM Cep in the literature span from 1939 until 2007, the time history is rather spotty, and there are few magnitudes before 2000. 

\citet{sic08} invoked several possible mechanisms to explain the large rapid variability of GM Cep's optical magnitude, the fast rotation rate, and the strong mid-IR excesses. The rapid variability can be explained by the strong outbursts of FUor systems (which brighten by $\geqslant$ 4 mag), in which the mass accretion rate through the circumstellar disk of a young star increases by orders of magnitude \citep{har96}. Another proto-stellar system, EXor (with outbursts $\geqslant$ 2 mag), was also interpreted as a mass accretion event \citep{leh95} and proposed to be similar to GM Cep. \citet{sic08} also give comparisons between the observational features of GM Cep and several better-known systems. For example, RW Aur, which is often quoted as a triple system, shares the features of a strong and variable P Cygni H$\alpha$ profile, a powerful disk, a large accretion rate, and a strong double-peaked $OI$ emission line with GM Cep \citep{ghe93, ale05, suy75}. Another similar system is GW Ori, a 1-Myr old G5 star with a fast rotation rate of $V\sin i$ = 43 km s$^{-1}$ \citep{bou86}, variability up to 1 mag in JHK\footnote{VizieR Online Data Catalog, II/250 \citep{sam04}}, and strong IR excess \citep{mat91, mat95}. CW Tau, a K3 star, has large magnitude variations of 2 mag\footnotemark[\value{footnote}] , a rapid rotation rate of $V\sin i$ = 28 km s$^{-1}$ \citep{muz98}, a P Cygni H$\alpha$ profile, and a deep, broad $OI$ absorption at 7773 \AA. McNeil's Nebula \citep{mcn04} has its emission line spectrum at optical wavelengths similar to the spectrum of GM Cep. KH 15D, a pre-main sequence binary system with a precessing disk or ring \citep{ham05}, is another system that provides an example of a possible explanation for the mechanism of GM Cep. However, without a long-term light curve of GM Cep, these physical explanations cannot be properly  tested, and the observational comparisons cannot be made. 

A long-term light curve can be used to search for outbursts, periodicities, repetitive features, and other observational features that these mechanisms predict. To obtain a long-term light curve, we visited Harvard College Observatory and Sonneberg Observatory, searched through the archival plates showing this field, and obtained 186 magnitude estimates from 1895 until 1993. We also collected the 75 visual observations from the database of the American Association of Variable Star Observations (AAVSO) from 2006 to present. A long-term light curve for GM Cep was plotted from these data.

\section{Data}

The majority of the world's archival photographic plates are now preserved at Harvard College Observatory (Cambridge, Massachusetts) and Sonneberg Observatory (Germany). The Harvard collection contains roughly 500,000 archival plates with complete sky coverage from mid-1880 to 1989, with a gap from 1953 to 1968. A large fraction of these plates are patrol plates, with a typical limiting magnitude (in the B band) of 12-15. There are also many series plates, with larger plate scale and deep limiting magnitudes ($\sim$15-18). The description of the patrol and series plates can be found on the HCO website\footnote{http://www.cfa.harvard.edu/hco/collect.html}. Most of the patrol plates are not deep enough to show GM Cep. As a result, our search focused on the series plates. Sonneberg Observatory was built in 1925 and has roughly 300,000 plates taken from the early 1930s until present, with patrol plates still ongoing. The magnitude limit of the series plates is $\sim$14-18, so many of these plates are deep enough to show GM Cep. The exposure times for the series plates range from $\sim$ 40 mins to 2 hrs. Most of the archival plates are in blue sensitivity, which closely matches the Johnson B band. Indeed, the Harvard plates provided the original definition of the B band, and the same spectral sensitivity is kept for the photoelectric and CCD magnitudes. With the comparison sequences measured in modern B magnitudes, the differential magnitudes from the old plates are now exactly in the Johnson B-magnitude system. 

Before looking through the plates, we set up our own comparison star sequence. The sequence was obtained at Sonoita Research Observatory\footnote{http://www.sonoitaobservatories.org/sonoita\_research\_observatory.html}, located near the town of Sonoita, AZ. The observatory has a 35cm (C14) Schmidt-Cassegrain telescope equipped with an SBIG STL-1000E CCD camera with Johnson-Cousins BVRI filters as well as a clear filter. The pixel scale of the telescope is 1.25 arcsec/pixel, with a 20$\times$20 arcmin field of view. All-sky photometry was obtained, using nightly standards \citep{lan83, lan92} on several photometric nights. The magnitudes and positions of the comparison stars are shown in Table 1.

We searched through all the series plates at Harvard and Sonneberg, and some of the patrol plates at Harvard (specifically, the Damon plates with a scale of 580"/mm and limiting magnitude 14-15 from years 1965 to 1990), and found 186 plates with images of GM Cep. All of these plates have blue sensitivity except for 10 Damon plates (6 DNY plates with visual sensitivity and 4 DNR plates with red sensitivity). We recorded all the plate numbers, dates, and the estimated GM Cep magnitudes. 

Each of the GM Cep magnitudes was obtained by taking the average of two or three independent estimations of the same plate. Our magnitude measurements were taken by visually examining each (back-illuminated) plate using a handheld loupe or microscope. Magnitudes were estimated by directly comparing the radius of GM Cep against the radii of nearby comparison stars. On photographic plates, only the objects with magnitude close to the limiting magnitude of the plates show a Gaussian profile. GM Cep is a relatively bright object, for which the central (Gaussian) portion of the star image is saturated. In this case, there is a sharp edge on the star profile, and human eyes are quite good at measuring the radius. The relation functions between the radii and the magnitudes are shown in \citet{sch91}. For our purpose here, as we are choosing comparison stars with comparable brightness on both sides (brighter and fainter) of GM Cep, the relation in such a small region can be approximated to be linear, and the uncertainty caused by the non-linear effect is much smaller than the measurement uncertainty itself. The field of GM Cep is not crowd at all, and all the measurements are well performed.

From our experience and the quantitative studies, our visual method is comparable in accuracy with methods based on two-dimensional scans of the plates and with the use of an Iris Diaphragm Photometer \citep{sch08, sch91, sch81, sch83b}. The measurement error on the magnitude estimation varies slightly among different plates. From the experience of the work on archival plates by our group at Louisiana State University, we can take a typical measurement error value of $\sim$ 0.15 mag \citep{sch83a, sch91, sch05, col09, pag09}. According to our data of GM Cep, the magnitude of each plate has been measured 2-3 times, and the average RMS of different measurements is 0.15, which provides us a typical measurement uncertainty of the magnitudes. In an archival plate study of nova QZ Aur \citep{xia10}, we calculate the standard deviation of the data points that are out of its eclipse. The standard deviation comes out to be 0.16 mag, which is compatible with the value we adopted here.

Table 2 records data from Sonneberg, and Table 3 records data from Harvard. Both of these tables are sorted in order of ascending time. The first column lists the plate number. The second and third columns show the date when the plate was taken, and the corresponding Julian day number. The fourth column lists our measured magnitudes. Our data show the long -term behavior of GM Cep from 1895 until 1993. The long-term light curve in the B band is plotted in Figure 1, and the light curve in the most densely sampled time interval (1935 to 1945) is plotted in Figure 2. B band data from \citet{sic08} are plotted on the same figure, which extends the time range to 2006-2007. 

The American Association of Variable Star Observers (AAVSO) has a substantial database of 75 V band magnitudes observed by two amateur astronomers. These data are available upon request at the AAVSO website\footnote{http://www.aavso.org}. The light curve from the combination of the AAVSO data, our V-band magnitudes from the DNY plates, and data from \citet{sic08} are shown in Figure 3. No measurement uncertainties are available for the AAVSO data, so we take 0.15 mag as a typical measurement error. Figure 4 displays the densely sampled V-band light curve from 2006 to 2009.

%We took photometric observations of GM Cep using the Robotic Optical Transient Search Experiment (ROTSE) telescope at McDonald Observatory, TX on May, 2009. ROTSE is a 0.45-m robotic reflecting telescope which is managed by a fully-automated system. The telescope has a $2^\circ$.64 diameter flat field of view. The limiting magnitude of the telescope is $\sim 17-18$ for a typical 20 second exposure time, which is sufficient for the observation of the bright object GM Cep \citep{ake03}. 

%Our data was taken on two separate nights, May 18th and 19th, 2009 UT. The exposure time of our observation is 20 seconds for each image. The total number of images we got is 392, with 40 taken on May 18, and 352 on May 19, and the time coverage is 3 hours for each night. No filter is used in our observation, and our broad-band magnitude covers $\sim 4000 - 9000 \AA$ wavelength, which corresponds to a combination of B, V, R, I bands in the Johnson-Cousins UBVRI photometric system. We chose a star A which has a comparable magnitude with GM Cep as the comparison star, and another star G as a check star. 

\section{Light Curve Analysis}

Different mechanisms have been proposed for the pre-main sequence star magnitude variations, including the rotation of a star with cool or hot spots, and the irregular UX-or stars \citep{her94}. However, none of them can explain the 2 - 2.5 mag variations within $\sim$10 days seen for GM Cep \citep{sic08}. Several comparable systems are pointed out in \citet{sic08} as possible explanations of the variation, e.g. FUors, EXors, RW Aur, GW Ori, CW Tau, McNeil's Nebula, and KH 15D. KH 15D was ruled out by \citet{sic08}, as it could not explain the high luminosity of GM Cep. All the remaining systems share some common features with GM Cep, as stated in Section 1. \citet{sic08} concluded that the variability mechanism is probably dominated by strong increases of the accretion rate. From our long-term light curve, we are able to analyze the behavior of GM Cep during the past century and compare it with all these listed possibilities.  

From our data and Figures 1 and 2, we see that the magnitude varies between 13.7 and 16.4, with most of the measures between 14.0 and 14.5. From the light curve between years 1938 and 1944, we see both rapid increases and decreases in magnitude (e.g. $\sim$1.1 mag increase from Aug. 2, 1938 to Aug. 19, 1938, $\sim$1 mag decrease from Jul. 25, 1941 to Sep. 15, 1941), which is in agreement with what \citet{sic08} found. The same rapid variation is found in AAVSO V band data, as shown in Figures 3 and 4. 

We also checked for periodicity in our data. Periodicity is expected if it is a binary system, with strong periodic modulations if the system is like KH 15D. Now that we have enough data, we can examine this possibility. We ran discrete Fourier transforms on both V band data from AAVSO and B band data from Harvard and Sonneberg. First, we constructed a smoothed light curve in both bands, which represents the long term variation behavior of GM Cep. By subtracting the smoothed light curve, we removed the long time scale variations and can search for the short timescale flickering that might be periodic. We ran a discrete Fourier transform analysis on both sets of data and found no significant period within the range 0.5 to 100 days. For the B band data from Harvard and Sonneberg, to get rid of the effect from long term variation of the light curve (especially the dips which are as large as 2 magnitudes), we picked a subsample of all the data points with magnitude between 13.75 and 14.75, which are not part of the dips. Another subsample we chose was the data between years 1935 and 1938, which is roughly constant before a dip, as shown in Figure 2. We ran the same discrete Fourier transforms on both subsamples, and neither of these show a significant period within the range 0.5 to 100 days. 

We made histograms of the magnitude distribution for both bands, which are shown in Figure 5. If the variability is caused by accretion, the light curve will have episodic outbursts (`shots') superposed on some quiescence state, and the corresponding magnitude histogram will show a cut-off at a higher magnitude and an extended tail to the lower magnitude. If the variations are caused by changing extinction, the light curve will have dips superposed on some quiescent state with roughly a constant magnitude, and the resulting magnitude histogram will be like a cut-off at a lower magnitude and an extended tail to the higher magnitude. From the figure we see a cut-off at the magnitude of $\sim$14, and a long extended tail to 16.5. We do not see how episodic flares from accretion can cause the system to spend most of its time in a nearly constant bright state. That is, with multiple shots (even if of some constant amplitude) superposed, the light curve should frequently be brighter than that of one shot due to the overlap of multiple shots, leading to a bright tail in the histogram. With the histogram showing a tail to the \textit{faint} side, we have an effective argument that the variation is not dominated by flares caused by accretion. 

Could the tail be caused by the detection thresholds of our plates? To test this, we checked all our data and sources. The detection threshold effect is most involved in the patrol plates from Harvard, i.e. the Damon plates in our data set. All the series plates at Sonneberg and Harvard are deep enough to obtain a GM Cep magnitude measurement. As a result, we made a magnitude histogram for data from Sonneberg series plates only and one for data exclusively from Harvard MC plates. These plots are shown in Figure 6. The Harvard MC histogram does not show any significant trend (although there are relatively few plates), while the Sonneberg histogram shows a cut-off at $\sim$14 mag and a significant extended tail, which is what we found above. As a result, we conclude that the faint tail in the histogram is not caused by threshold effects on the plates.

We are now able to compare GM Cep with the possible mechanisms listed previously. The most obvious property of the light curve is that GM Cep has not undergone any substantial outburst since 1895, and the light curve itself indicate that the variability is due to dips caused by extinction instead of outbursts caused by accretion. The conclusion is confirmed by the magnitude histogram in both B and V bands. FUor stars are characterized by large outbursts with typical rises of 5 magnitudes in a year or so \citep{har96}. In EXor stars, recurrent bursts with amplitudes $> 2$ mag, which last $\leqslant 1$ yr, are found \citep{her01}. Given our sampling time and sensitivity, similar features in GM Cep could not be missed. Thus, our long-term data rule out the associations with FUor or EXor systems. McNeil's Nebula \citep{mcn04}, which shows outbursts with EXor or FUor type eruptions, can also be excluded by our light curve. Our data are compatible with more evolved T Tauri systems,  whose magnitude variability is typically $\leqslant 0.4$ mag, with no significant changes over many years  \citep{gra07, gra08}. T Tauri systems are also likely populating an old cluster such as Tr37. However, the T Tauri system is not able to produce the huge dips ($\sim 2$ mags) in the light curves. Certain unusual T Tauri stars were found to share spectral features with GM Cep \citep{sic08}. These systems are RW Aur, GW Ori, and CW Tau, and we compared the light curves of these systems to that of GM Cep. RW Aur shows variations of $> 2$ mags, but no dips was found in the long-term light curve from both the literature \citep{ahn57, bec01, pet01} and AAVSO data. GW Ori showes variations of $< 0.2$ mags, with eclipses of $\sim 0.4$ mags \citep{she92, she98}, which is not compatible with GM Cep. For CW Tau, only a few data points were obtained from literature, and no variation was found in the long-term photometric monitoring \citep{wal99}. As a result, none of the light curves of these three systems are compatible with GM Cep. The absence of periodic features in the light curve also excludes the possibility of GM Cep being a KH 15D type star.

\section{Summary}

In this paper, we present data of GM Cep from all available series archival plates and some patrol plates from Harvard College Observatory and Sonneberg Observatory. We obtained 186 new magnitudes for GM Cep (176 in blue, 6 in visible, and 4 in red) ranging from 1895 until 1993. Another 75 V band magnitudes were drawn from the AAVSO database. By combining our data from archival plates, AAVSO data, and previously-published data collected by \citet{sic08}, long term B and V band light curves for GM Cep were constructed. The B band light curve shows a generally constant magnitude ($\sim$14-14.5) with occasional dips to $\sim$16.5. Fast variations are found in both the B and V band light curves. 

The magnitude histograms in both B and V bands show cut-offs at the low magnitude (bright) end and long extended tails at the high magnitude (faint) end, which implies that the light curve is composed of dips caused by varying extinction instead of outbursts caused by accretion superposed on quiescence state. The lack of large outbursts in the past century implies that it is not a FUor or EXor star, or a McNeil's Nebula type star. The lack of periodicity in the light curve also excludes the possibility of GM Cep being a KH 15D type star. Several special cases of T Tauri stars (RW Aur, GW Ori and CW Tau) were checked, but none of these light curves are compatible with that of GM Cep.

\acknowledgements
We thank the many observers and curators for the archival plate collections at Harvard College Observatory and at Sonneberg Observatory.  The work in this paper would not be possible without their patient and hopeful work over the last century. We would like to thank Bradley Schaefer, Ashley Pagnotta and Andrew Collazzi for their help and useful discussions. We also thank the amateur astronomers of the AAVSO for providing the V magnitude values of GM Cep in 2006-2008.

\clearpage

\begin{figure}
\epsscale{.80}
\plotone{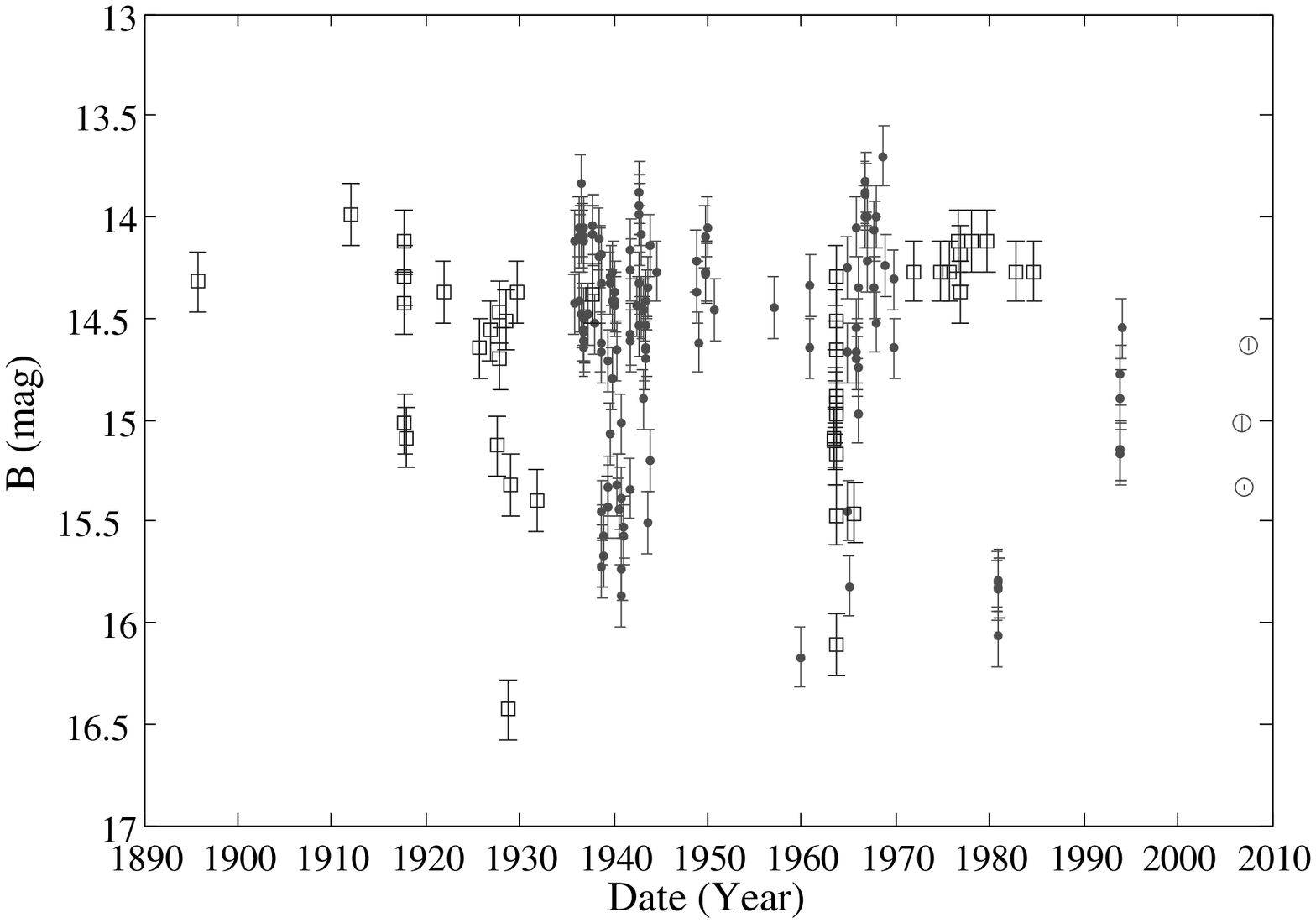}
\caption{Long term light curve of variable star GM Cep from Harvard College Observatory plates (empty squares), Sonneberg Observatory plates (filled circles), and \citet{sic08} data (empty circles). All these data are in blue band. The archival data spans from 1895 up until 1993, with both being consistent with each other. The \citet{sic08} data extend the light curve with blue magnitudes in 2006-2007. }
\end{figure}

\begin{figure}
\epsscale{.80}
\plotone{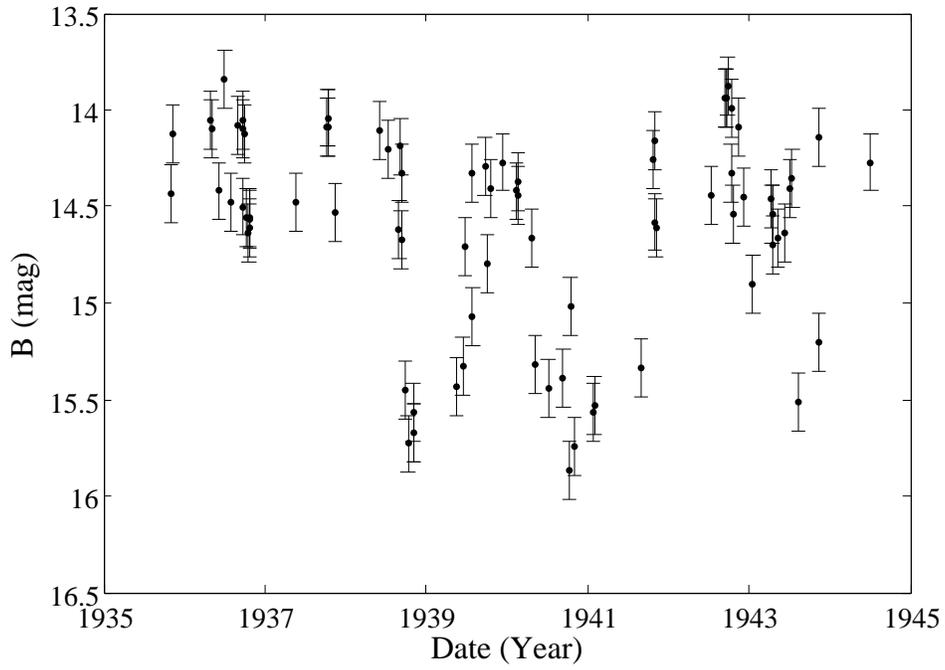}
\caption{GM Cep light curve between years 1935 and 1945. This is the most densely sampled time interval. From this light curve we see significant magnitude variation, with repetitive dips up to $\sim$2 mags. }
\end{figure}

\begin{figure}
\epsscale{0.80}
\plotone{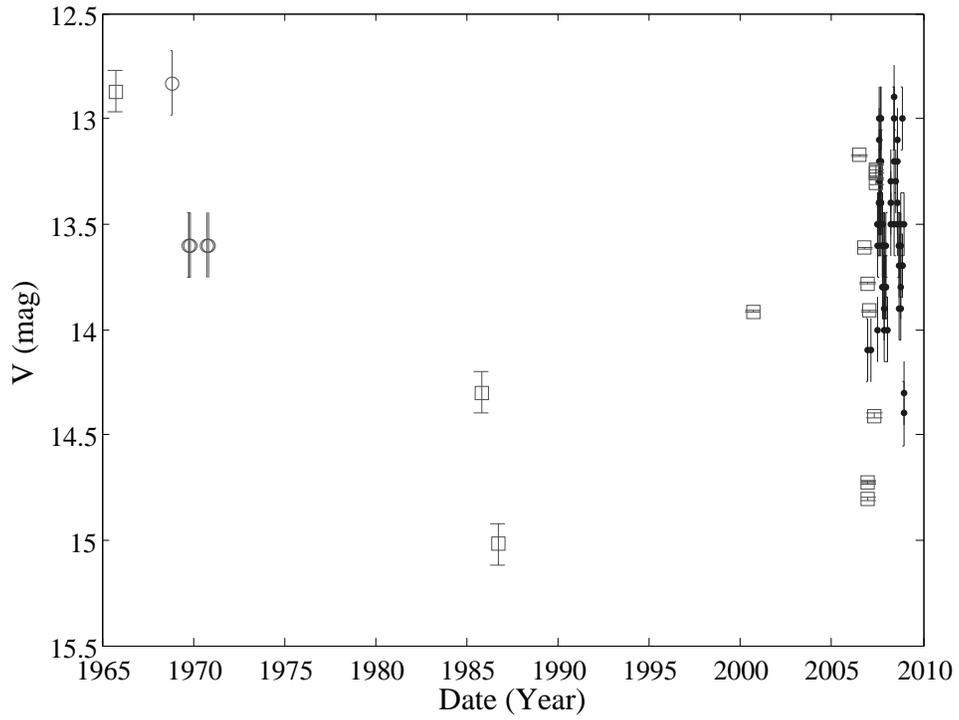}
\caption{V band GM Cep light curve from \citet{sic08}, AAVSO data, and our data from DNY plates. The filled circles are AAVSO data from 2006 to 2008, the empty squares are \citet{sic08} data from 1965, and the empty circles are DNY plates data in 1968-1970. The data shows a $\sim$2 mag magnitude variability in V band. }
\end{figure}

\begin{figure}
\epsscale{0.80}
\plotone{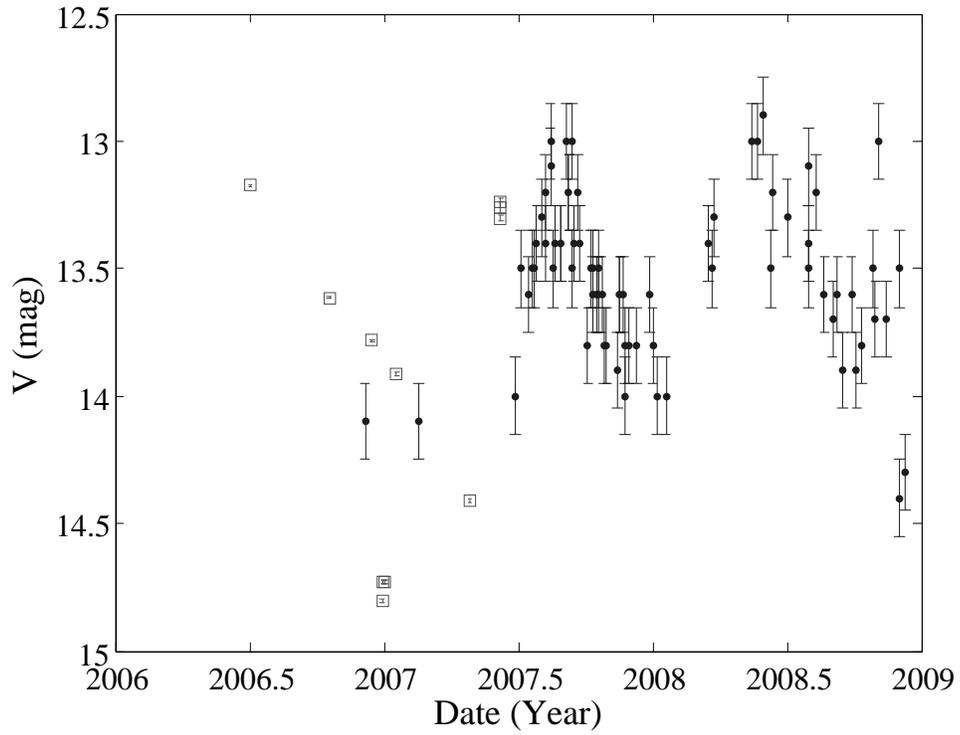}
\caption{V band GM Cep light curve from \citet{sic08} and AAVSO data in years 2006 to 2008. The filled circles are AAVSO data, and the empty squares are data from \citet{sic08}. These two datasets are consistent and show rapid variability with a range of $\sim$ 2 mag. }
\end{figure}

\begin{figure}
\epsscale{1.10}
\plottwo{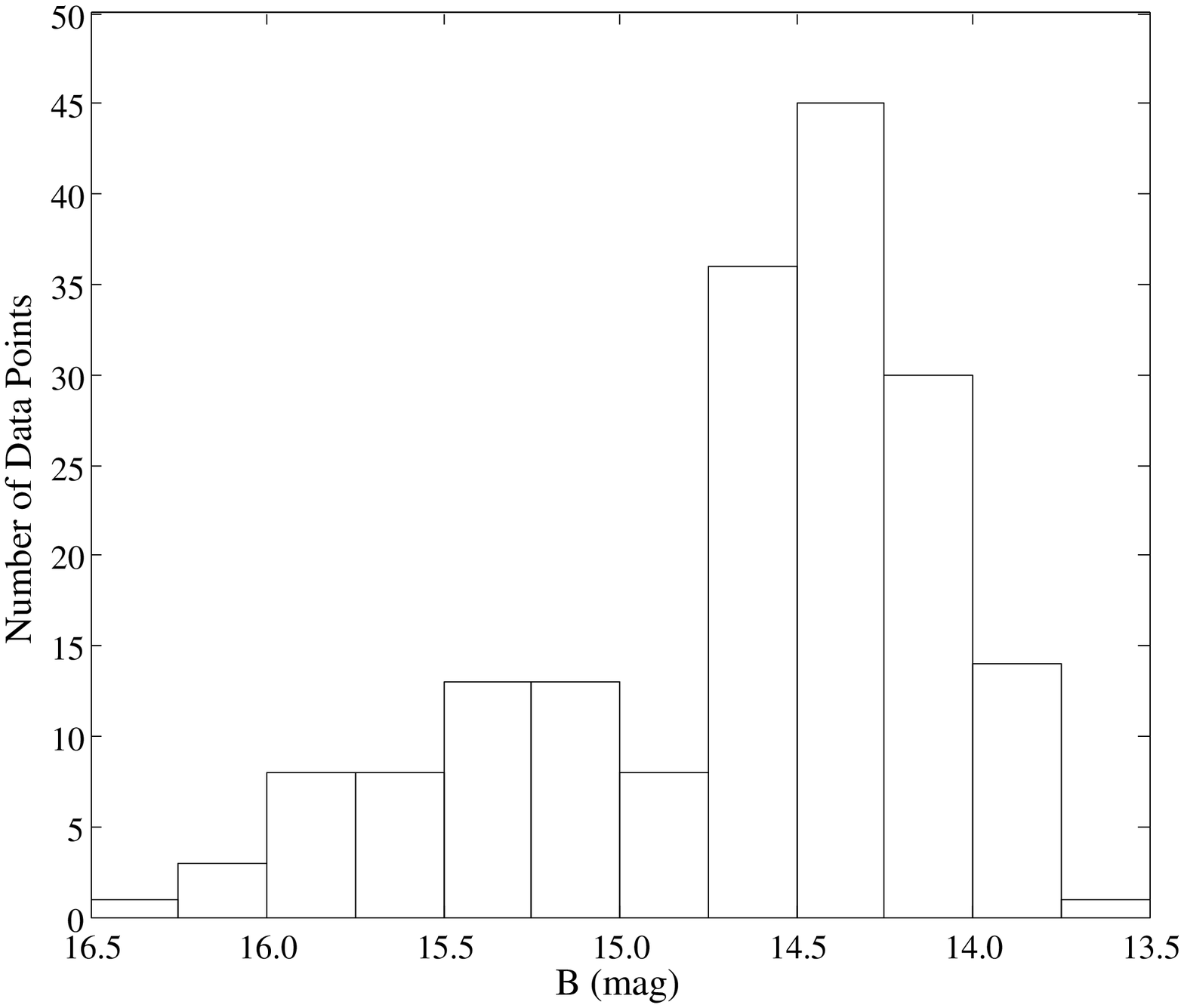}{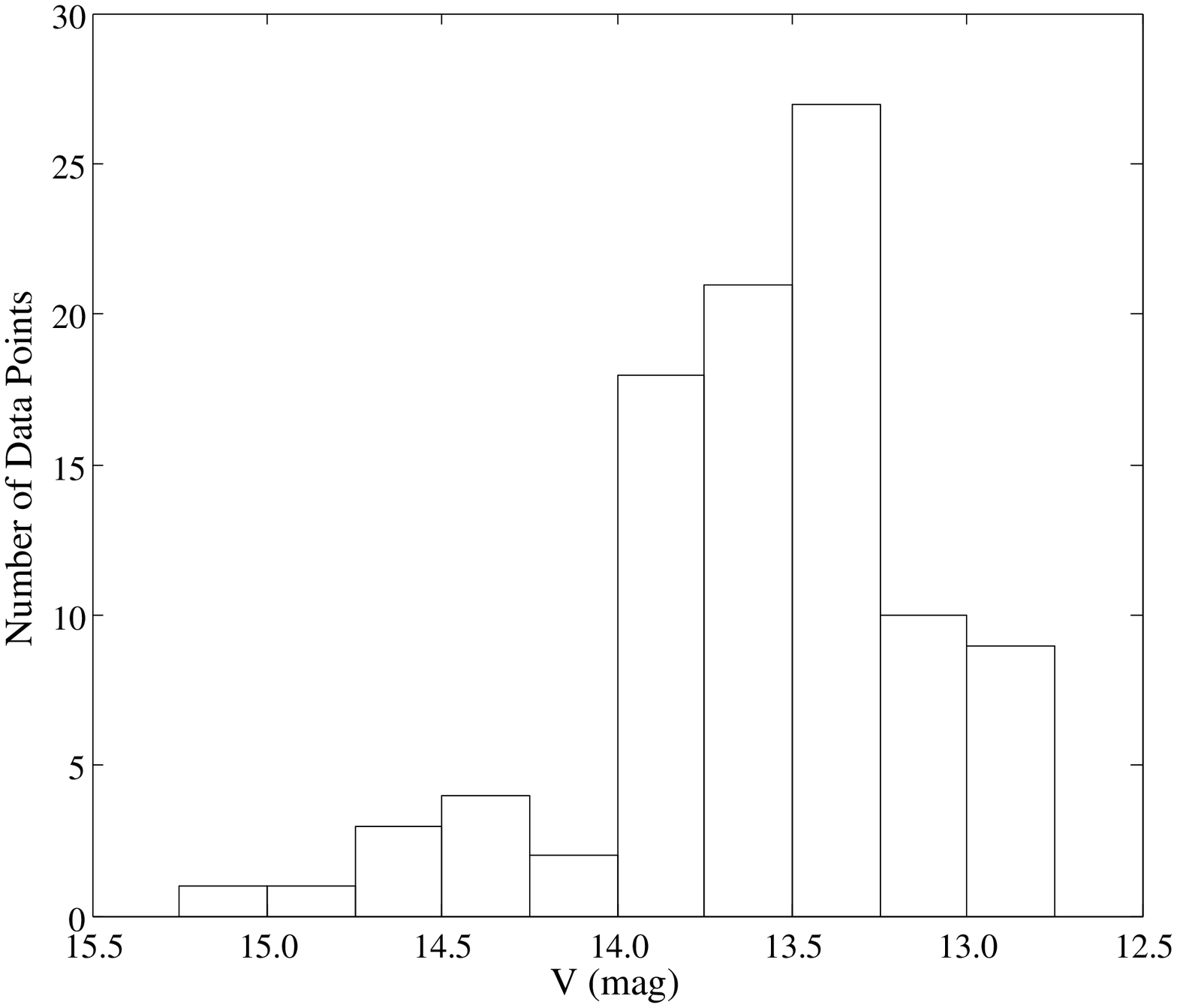}
\caption{Distribution of magnitudes. Left panel: B band magnitude distribution from Harvard and Sonneberg plates and 4 B band data from \citet{sic08}. Right panel: V band magnitude distribution from AAVSO data and 16 V band data from \citet{sic08}. Both of these distributions show long extended tails to the high magnitude (faint) region. }
\end{figure}

\begin{figure}
\epsscale{1.00}
\plottwo{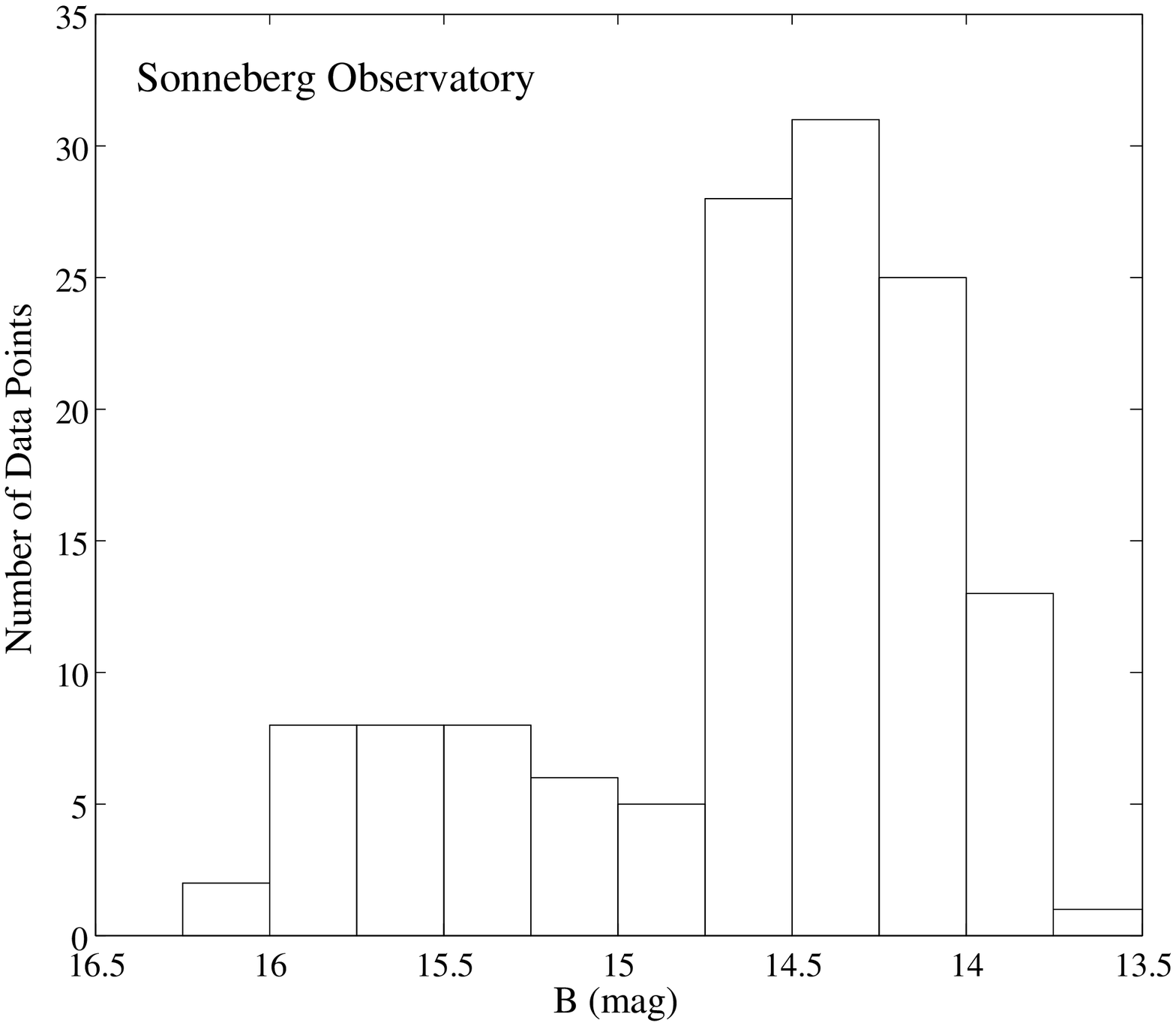}{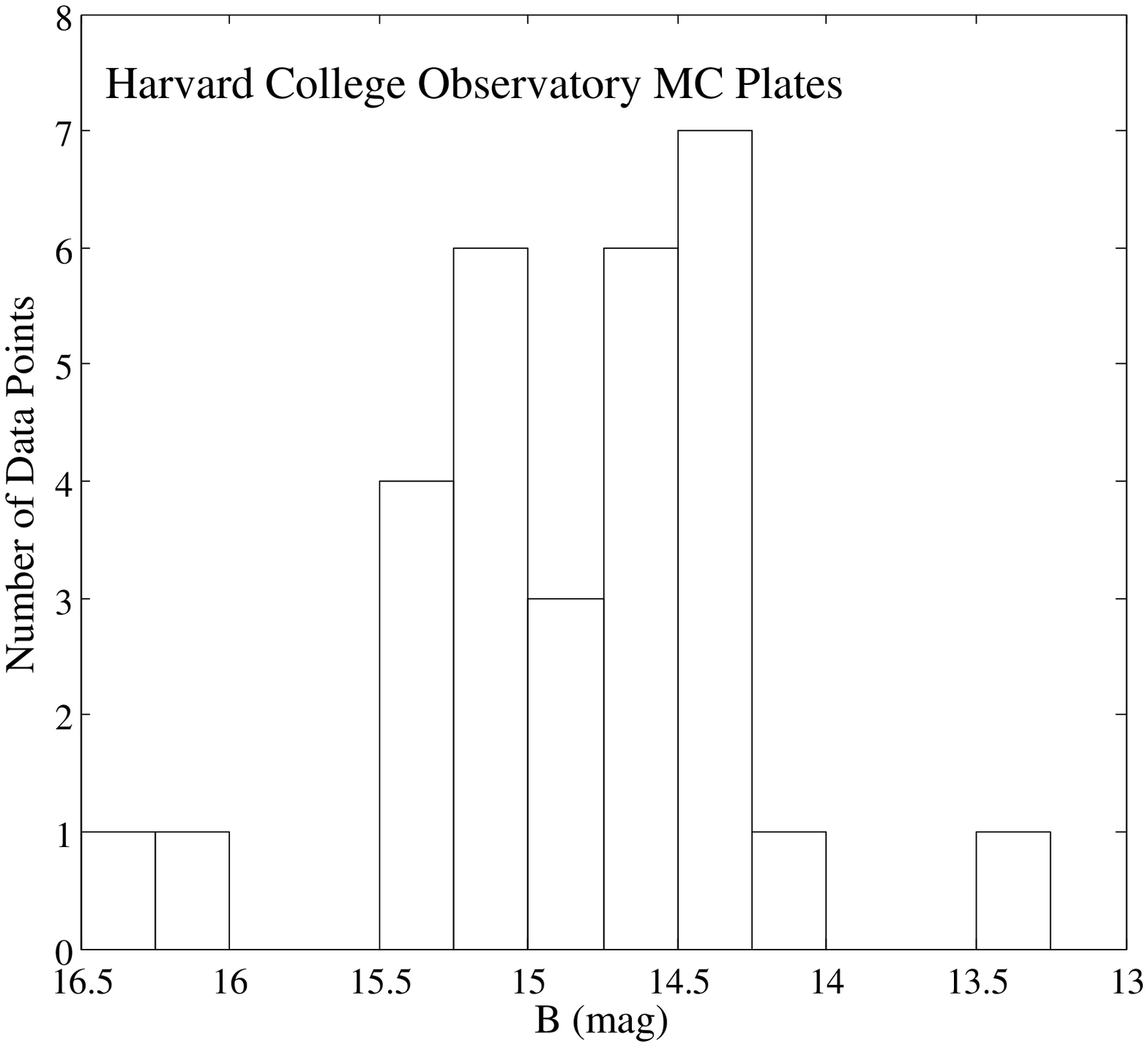}
\caption{Distribution of magnitudes. Left panel: B band magnitude distribution from Sonneberg plates. Right panel: B band magnitude distribution from Harvard College Observatory MC plates. Both of these two series are deep plates which have limiting magnitudes fainter than 16.5 mag. As such, no detection threshold effect is involved in these two histograms. The long extended tail is significant in Sonneberg plate histogram. }
\end{figure}

\clearpage

\begin{deluxetable}{lllllll}
\tabletypesize{\scriptsize}
%\rotate
\tablecaption{Comparison Sequence for GM Cep}
\tablewidth{0pt}
\tablehead{
\colhead{Field} &
\colhead{Star} & 
\colhead{RA (J2000)} & 
\colhead{Dec (J2000)} &
\colhead{B (mag)} &
\colhead{V (mag)} &
\colhead{R (mag)}
}
\startdata
GM Cep	&	A	&	324.593662	&	57.536787	&	16.313	&	14.213	&	12.984	\\
GM Cep	&	B	&	324.529226	&	57.508117	&	16.015	&	14.961	&	14.364	\\
GM Cep	&	C	&	324.563184	&	57.492816	&	15.445	&	14.837	&	14.455	\\
GM Cep	&	D	&	324.543391	&	57.505287	&	15.333	&	14.357	&	13.770	\\
GM Cep	&	E	&	324.580052	&	57.532424	&	14.628	&	13.601	&	13.187	\\
GM Cep	&	F	&	324.586443	&	57.487231	&	14.389	&	13.358	&	12.770	\\
GM Cep	&	G	&	324.600939	&	57.556202	&	13.374	&	12.829	&	12.513	\\

\enddata
%% Text for table notes should follow after the \enddata but before
%% the \end{deluxetable}. Make sure there is at least one \tablenotemark
%% in the table for each \tablenotetext.
%\tablecomments{Table \ref{tbl-1} is published in its entirety in the 
%electronic edition of the {\it Astrophysical Journal}.  A portion is 
%shown here for guidance regarding its form and content.}
%\tablenotetext{a}{Sample footnote for table~\ref{tbl-1} that was generated
%with the deluxetable environment}
%\tablenotetext{b}{Another sample footnote for table~\ref{tbl-1}}
\end{deluxetable}

%% If you use the table environment, please indicate horizontal rules using
%% Do not put multiple tabular environments within a single table.
%% The optional \label should appear inside the \caption command.

\clearpage

\begin{deluxetable}{llll}
\tabletypesize{\scriptsize}
%\rotate
\tablecaption{Magnitude of GM Cep from Sonneberg Observatory Plates}
\tablewidth{0pt}
\tablehead{
\colhead{Plate Number} &
\colhead{Date (dd-mmm-yy)} & 
\colhead{Julian Day Number} & 
\colhead{Magnitude} 
}
\startdata	
F 1799	&	23-Sep-1935	&	2428069	&	14.43	\\
F 1795	&	2-Oct-1935	&	2428078	&	14.12	\\
F 1867	&	18-Mar-1936	&	2428246	&	14.05	\\
F 1891	&	25-Mar-1936	&	2428253	&	14.10	\\
F 1901	&	26-Apr-1936	&	2428285	&	14.42	\\
F 1919	&	25-May-1936	&	2428314	&	13.84	\\
F 1938	&	24-Jun-1936	&	2428344	&	14.48	\\
F 1947	&	20-Jul-1936	&	2428370	&	14.08	\\
F 1955	&	14-Aug-1936	&	2428395	&	14.50	\\
F 1961	&	16-Aug-1936	&	2428397	&	14.10	\\
F 1960	&	16-Aug-1936	&	2428397	&	14.05	\\
F 1970	&	23-Aug-1936	&	2428404	&	14.12	\\
F 1769	&	1-Sep-1936	&	2428413	&	14.56	\\
F 1982	&	10-Sep-1936	&	2428422	&	14.64	\\
F 1993	&	13-Sep-1936	&	2428425	&	14.56	\\
F 1997	&	15-Sep-1936	&	2428427	&	14.57	\\
F 2001	&	16-Sep-1936	&	2428428	&	14.61	\\
F 2118	&	12-Apr-1937	&	2428636	&	14.48	\\
F 2190	&	1-Sep-1937	&	2428778	&	14.09	\\
F 2193	&	2-Sep-1937	&	2428779	&	14.04	\\
F 2200	&	6-Sep-1937	&	2428783	&	14.04	\\
F 2203	&	7-Sep-1937	&	2428784	&	14.09	\\
F 2213	&	6-Oct-1937	&	2428813	&	14.53	\\
F 2292	&	25-Apr-1938	&	2429014	&	14.11	\\
F 2310	&	1-Jun-1938	&	2429051	&	14.20	\\
F 2327	&	23-Jul-1938	&	2429103	&	14.62	\\
A 2340	&	31-Jul-1938	&	2429111	&	14.19	\\
F 2342	&	1-Aug-1938	&	2429112	&	14.67	\\
F 2344	&	2-Aug-1938	&	2429113	&	14.33	\\
F 2351	&	19-Aug-1938	&	2429130	&	15.45	\\
F 2359	&	1-Sep-1938	&	2429143	&	15.73	\\
F 2387	&	24-Sep-1938	&	2429166	&	15.57	\\
F 2393	&	26-Sep-1938	&	2429168	&	15.67	\\
F 2397	&	27-Sep-1938	&	2429169	&	15.67	\\
F 2476	&	11-Apr-1939	&	2429365	&	15.43	\\
F 2495	&	6-May-1939	&	2429390	&	15.33	\\
F 2497	&	17-May-1939	&	2429401	&	14.71	\\
F 2509	&	16-Jun-1939	&	2429431	&	14.33	\\
F 2522	&	21-Jun-1939	&	2429436	&	15.07	\\
F 2529	&	15-Aug-1939	&	2429491	&	14.29	\\
F 2546	&	24-Aug-1939	&	2429500	&	14.80	\\
F 2551	&	8-Sep-1939	&	2429515	&	14.41	\\
F 2567	&	2-Nov-1939	&	2429570	&	14.27	\\
F 2616	&	6-Jan-1940	&	2429635	&	14.42	\\
F 2626	&	10-Jan-1940	&	2429639	&	14.37	\\
F 2638	&	12-Jan-1940	&	2429641	&	14.44	\\
F 2674	&	16-Mar-1940	&	2429705	&	14.66	\\
F 2679	&	1-Apr-1940	&	2429721	&	15.32	\\
F 2703	&	27-May-1940	&	2429777	&	15.44	\\
F 2726	&	4-Aug-1940	&	2429846	&	15.39	\\
F 2735	&	3-Sep-1940	&	2429876	&	15.87	\\
F 2741	&	5-Sep-1940	&	2429878	&	15.02	\\
F 2748	&	25-Sep-1940	&	2429898	&	15.74	\\
F 2788	&	20-Dec-1940	&	2429984	&	15.57	\\
F 2794	&	22-Dec-1940	&	2429986	&	15.53	\\
F 2865	&	25-Jul-1941	&	2430201	&	15.34	\\
F 2879	&	15-Sep-1941	&	2430253	&	14.26	\\
F 2885	&	20-Sep-1941	&	2430258	&	14.58	\\
F 2897	&	23-Sep-1941	&	2430261	&	14.16	\\
F 2908	&	27-Sep-1941	&	2430265	&	14.61	\\
F 3035	&	5-Jun-1942	&	2430516	&	14.44	\\
F 3048	&	6-Aug-1942	&	2430578	&	13.94	\\
F 3051	&	10-Aug-1942	&	2430582	&	13.94	\\
F 3056	&	15-Aug-1942	&	2430587	&	13.94	\\
F 3062	&	18-Aug-1942	&	2430590	&	13.88	\\
F 3067	&	2-Sep-1942	&	2430605	&	14.33	\\
F 3071	&	5-Sep-1942	&	2430608	&	13.99	\\
F 3074	&	10-Sep-1942	&	2430613	&	14.54	\\
F 3092	&	5-Oct-1942	&	2430638	&	14.09	\\
F 3098	&	28-Oct-1942	&	2430661	&	14.45	\\
F 3112	&	11-Dec-1942	&	2430705	&	14.90	\\
F 3159	&	3-Mar-1943	&	2430787	&	14.46	\\
F 3169	&	7-Mar-1943	&	2430791	&	14.54	\\
F 3177	&	9-Mar-1943	&	2430793	&	14.54	\\
F 3180	&	10-Mar-1943	&	2430794	&	14.70	\\
F 3194	&	5-Apr-1943	&	2430820	&	14.66	\\
F 3201	&	2-May-1943	&	2430847	&	14.64	\\
F 3207	&	28-May-1943	&	2430873	&	14.41	\\
F 3210	&	7-Jun-1943	&	2430883	&	14.35	\\
F 3214	&	2-Jul-1943	&	2430908	&	15.51	\\
F 3087	&	3-Oct-1943	&	2431001	&	14.14	\\
F 3269	&	6-Oct-1943	&	2431004	&	15.20	\\
F 3336	&	28-May-1944	&	2431239	&	14.27	\\
F 3673	&	9-Sep-1948	&	2432804	&	14.22	\\
F 3682	&	2-Oct-1948	&	2432827	&	14.37	\\
F 3696	&	27-Nov-1948	&	2432883	&	14.62	\\
F 3760	&	20-Aug-1949	&	2433149	&	14.10	\\
F 3778	&	20-Sep-1949	&	2433180	&	14.27	\\
F 3792	&	24-Sep-1949	&	2433184	&	14.28	\\
F 3789	&	26-Sep-1949	&	2433186	&	14.27	\\
F 3800	&	20-Oct-1949	&	2433210	&	14.05	\\
F 3884	&	14-Aug-1950	&	2433508	&	14.46	\\
F 4564	&	22-Nov-1956	&	2435800	&	14.45	\\
F 4933	&	11-Sep-1959	&	2436823	&	16.17	\\
F 5069	&	29-Sep-1960	&	2437207	&	14.65	\\
F 5077	&	14-Oct-1960	&	2437222	&	14.34	\\
F 5664	&	14-Sep-1964	&	2438653	&	14.25	\\
F 5676	&	3-Oct-1964	&	2438672	&	14.67	\\
F 5691	&	8-Nov-1964	&	2438708	&	15.45	\\
F 5704	&	10-Dec-1964	&	2438740	&	15.82	\\
F 5763	&	23-Aug-1965	&	2438996	&	14.05	\\
F 5771	&	21-Sep-1965	&	2439025	&	14.70	\\
F 5776	&	23-Sep-1965	&	2439027	&	14.55	\\
F 5790	&	16-Oct-1965	&	2439050	&	14.67	\\
F 5799	&	22-Oct-1965	&	2439056	&	14.35	\\
F 5813	&	16-Nov-1965	&	2439081	&	14.74	\\
F 5815	&	23-Nov-1965	&	2439088	&	14.97	\\
F 5887	&	13-Aug-1966	&	2439351	&	14.00	\\
F 5891	&	17-Aug-1966	&	2439355	&	13.88	\\
F 5897	&	10-Sep-1966	&	2439379	&	13.89	\\
F 5908	&	19-Sep-1966	&	2439388	&	13.83	\\
F 5913	&	21-Sep-1966	&	2439390	&	14.00	\\
F 5919	&	7-Oct-1966	&	2439406	&	14.00	\\
F 5926	&	6-Nov-1966	&	2439436	&	14.22	\\
F 6027	&	5-Aug-1967	&	2439708	&	14.35	\\
F 6029	&	8-Aug-1967	&	2439711	&	14.07	\\
F 6039	&	28-Sep-1967	&	2439762	&	14.52	\\
F 6057	&	20-Nov-1967	&	2439815	&	14.00	\\
F 6140	&	29-Jul-1968	&	2440067	&	13.70	\\
F 6171	&	11-Nov-1968	&	2440172	&	14.24	\\
F 6230	&	8-Sep-1969	&	2440473	&	14.31	\\
F 6238	&	5-Oct-1969	&	2440500	&	14.65	\\
SC 4450	&	7-Sep-1980	&	2444490	&	15.80	\\
SC 4455	&	16-Sep-1980	&	2444499	&	15.84	\\
SC 4466	&	3-Oct-1980	&	2444516	&	16.07	\\
SC 4469	&	10-Oct-1980	&	2444523	&	15.79	\\
SC 4470	&	10-Oct-1980	&	2444523	&	15.79	\\
SC 4473	&	28-Oct-1980	&	2444541	&	15.83	\\
SC 4474	&	1-Nov-1980	&	2444545	&	15.83	\\
GC 10944	&	1-Nov-1993	&	2449293	&	14.78	\\
GC 10943	&	1-Nov-1993	&	2449293	&	15.15	\\
GC 10945	&	2-Nov-1993	&	2449294	&	15.17	\\
GC 10946	&	3-Nov-1993	&	2449295	&	15.15	\\
GC 10949	&	Nov, 1993	&	2449296	&	14.90	\\
GC 10955	&	18-Nov-1993	&	2449310	&	14.55	\\
\enddata
%\tablenotetext{a}{The hand written of the plate number is blurred. It is not affecting our data and result, because the recorded time and date can be easily read, and the plate itself is not damaged. }
%% Text for table notes should follow after the \enddata but before
%% the \end{deluxetable}. Make sure there is at least one \tablenotemark
%% in the table for each \tablenotetext.
%\tablecomments{Table \ref{tbl-1} is published in its entirety in the 
%electronic edition of the {\it Astrophysical Journal}.  A portion is 
%shown here for guidance regarding its form and content.}
%\tablenotetext{a}{Sample footnote for table~\ref{tbl-1} that was generated
%with the deluxetable environment}
%\tablenotetext{b}{Another sample footnote for table~\ref{tbl-1}}
\end{deluxetable}

\begin{deluxetable}{llll}
\tabletypesize{\scriptsize}
%\rotate
\tablecaption{Magnitude of GM Cep from Harvard College Observatory Plates}
\tablewidth{0pt}
\tablehead{
\colhead{Plate Number} &
\colhead{Date (dd-mmm-yy)} & 
\colhead{Julian Day Number} & 
\colhead{Magnitude} 
}
\startdata
A 1580	&	20-Aug, 1895	&	2413426	&	14.32	\\
MC 1489	&	23-Dec-1911	&	2419394	&	13.99	\\
MC 13056	&	26-Jul-1917	&	2421436	&	14.43	\\
MC 13167	&	12-Aug-1917	&	2421453	&	14.12	\\
MC 13321	&	6-Sep-1917	&	2421478	&	15.02	\\
MC 13444	&	12-Sep-1917	&	2421484	&	14.29	\\
MC 14307	&	13-Nov-1917	&	2421546	&	15.09	\\
MC 17954	&	27-Sep-1921	&	2422960	&	14.37	\\
MC 21599	&	23-Sep-1925	&	2424417	&	14.65	\\
MC 22094	&	14-Sep-1926	&	2424773	&	14.56	\\
MC 22606	&	24-Jul-1927	&	2425086	&	15.13	\\
MC 22667	&	5-Sep-1927	&	2425129	&	14.47	\\
MC 22835	&	26-Oct-1927	&	2425180	&	14.70	\\
MC 23452	&	12-Jun-1928	&	2425410	&	14.51	\\
MC 23648	&	22-Sep-1928	&	2425512	&	16.43	\\
MC 23816	&	13-Nov-1928	&	2425564	&	15.32	\\
MC 24365	&	16-Jul-1929	&	2425809	&	14.37	\\
MC 25591	&	19-Aug-1931	&	2426573	&	15.40	\\
MC 29083	&	8-Sep-1937	&	2428785	&	14.38	\\
MC 38812	&	13-Jun-1963	&	2438194	&	15.09	\\
MC 38813	&	13-Jun-1963	&	2438194	&	15.10	\\
MC 38823	&	17-Jun-1963	&	2438198	&	15.47	\\
MC 38829	&	18-Jun-1963	&	2438199	&	14.51	\\
MC 38830	&	18-Jun-1963	&	2438199	&	14.97	\\
MC 38828	&	18-Jun-1963	&	2438199	&	14.92	\\
MC 38834	&	19-Jun-1963	&	2438200	&	14.29	\\
MC 38835	&	19-Jun-1963	&	2438200	&	15.17	\\
MC 38838	&	21-Jun-1963	&	2438202	&	14.89	\\
MC 38840	&	21-Jun-1963	&	2438202	&	14.66	\\
MC 38899	&	10-Aug-1963	&	2438252	&	16.11	\\
MC 38857	&	13-Jul-1965	&	2438955	&	15.46	\\
DNY 131		&	19-Sep-1968	&	2440119	&	12.83	\\
DNY 189		&	19-Aug-1969	&	2440453	&	13.60	\\
DNY 191		&	12-Sep-1969	&	2440477	&	13.60	\\
DNY 196		&	9-Oct-1969	&	2440504	&	13.60	\\
DNY 235		&	1-Sep-1970	&	2440831	&	13.60	\\
DNY 238		&	29-Sep-1970	&	2440859	&	13.60	\\
DNR 246		&	29-Oct-1970	&	2440889	&	12.79	\\
DNR 252		&	17-Nov-1970	&	2440908	&	12.75	\\
DNR 285		&	23-Aug-1971	&	2441187	&	12.79	\\
DNR 286		&	24-Aug-1971	&	2441188	&	13.01	\\
DNB 380		&	19-Oct-1971	&	2441244	&	14.27	\\
DNB 750		&	12-Aug-1974	&	2442272	&	14.27	\\
DNB 1112	&	3-Sep-1975	&	2442659	&	14.27	\\
DNB 1455	&	24-Jul-1976	&	2442984	&	14.12	\\
DNB 1520	&	22-Sep-1976	&	2443044	&	14.37	\\
DNB 1558	&	19-Oct-1976	&	2443071	&	14.19	\\
DNB 1894	&	2-Nov-1977	&	2443450	&	14.12	\\
DNB 2359	&	18-Aug-1979	&	2444104	&	14.12	\\
DNB 3218	&	26-Aug-1982	&	2445208	&	14.27	\\
DNB 4299	&	27-Aug-1984	&	2445940	&	14.27	\\
\enddata
%% Text for table notes should follow after the \enddata but before
%% the \end{deluxetable}. Make sure there is at least one \tablenotemark
%% in the table for each \tablenotetext.
%\tablecomments{Table \ref{tbl-1} is published in its entirety in the 
%electronic edition of the {\it Astrophysical Journal}.  A portion is 
%shown here for guidance regarding its form and content.}
%\tablenotetext{a}{Sample footnote for table~\ref{tbl-1} that was generated
%with the deluxetable environment}
%\tablenotetext{b}{Another sample footnote for table~\ref{tbl-1}}
\end{deluxetable}
\clearpage

\end{document}